\documentclass{optica-article}

\journal{opticajournal} 

\articletype{Research Article}

\usepackage{lineno}
\usepackage{siunitx}

\begin{document}

\title{Optically Writable Atomic Vapor Memory as a Substrate for Optical Reservoir Computing}

\author{Elizabeth Robertson\authormark{1, $\dagger$}, Mingwei Yang \authormark{2, $\dagger$}, Lina Jaurigue,\authormark{3} Guillermo Gallego~\authormark{4,5}, Kathy Lüdge,\authormark{3} and Janik Wolters \authormark{1,2,5,6,*}}

\address{\authormark{1}Institute of Space Research, German Aerospace Center (DLR), Berlin, Germany\\
\authormark{2} Institute of Physics, Technische Universit\"at Berlin, Berlin, Germany \\
\authormark{3} Institut für Physik, Technische Universität Illmenau, Illmenau, Germany\\
\authormark{4} Department of Electrical Engineering and Computer Science, Technical University of Berlin, Berlin, Germany\\
\authormark{5} Einstein Center Digital Future (ECDF), Berlin, Germany \\
\authormark{6} AQLS UG Haftungsbeschränkt, Germany \\
\authormark{$\dagger$} Both authors contributed equally.
}

\email{\authormark{*}janik.wolters@dlr.de} 

\newcommand{\ORAM}{ORAM}
\newcommand{\RC}{RC}
\newcommand{\AOM}{AOM}
\newcommand{\LMC}{MC}
\newcommand{\KR}{KR}
\newcommand{\BER}{BER}
\newcommand{\TBP}{TBP}
\newcommand{\NMSE}{NMSE}
\newcommand{\ELM}{ELM}
\newcommand{\Cs}{Cs}
\DeclareSIUnit\torr{Torr}

\DeclareFontFamily{U}{stix2bb}{}
\DeclareFontShape{U}{stix2bb}{m}{n} {<-> stix2-mathbb}{}

\NewDocumentCommand{\indicator}{}{\text{\usefont{U}{stix2bb}{m}{n}1}}


\begin{abstract*} 
We present an optical random access memory (ORAM) based on warm cesium (Cs) atomic vapor and demonstrate its operation as the physical substrate of a reservoir computer. Information is stored in the hyperfine population distribution of a Cs ensemble via optical pumping and retrieved through differential probe absorption. Spatial multiplexing via acousto-optic deflection provides eight addressable memory rails able to store up to 3.8 bits of information per rail. Employing this platform as a temporally multiplexed reservoir, we achieve a kernel rank (KR~$= 8.8 \pm 0.4$), and a minimum bit error rate of $0.02 \pm 0.01$ on the Exclusive-or (XOR) benchmark. We find the limited memory lifetime constrains the achievable temporal depth, encouraging further research into fast addressable memories. This constitutes the first demonstration of a free-space, optically writable atomic RAM as a substrate in an optical reservoir computing system.
\end{abstract*}


\section{Introduction}

Memory is a fundamental element of information processing, as evidenced by the central role of the read-write tape in the Turing machine’s definition of classical computability~\cite{turingComputableNumbersApplication1937}. 
Historically, encoding information in classical optical signals has proven to be a highly viable storage method. Optical Read-Only Memories (OROMs)---spanning analog formats like paper and photographs to robust digital media such as CDs, DVDs, and Blu-ray discs---are mature, commercialized technologies. 
Moreover, the high degree of parallelism offered by optical memory modules~\cite{mumbruOpticalMemoryComputing1999} continues to drive research into high-volume archival storage for data centers~\cite{chenEncodingDecodingMultidimensional2026, cheriereHolographicStorageCloud2025}.
In contrast, optical random access memories (ORAM) have not seen the same commercial success. 
This is primarily due to the overwhelming dominance of electronic counterparts (e.g., SRAM, DRAM, and flash memory), which provide versatile, multi-timescale storage natively integrated on silicon chips. 
However, as the computing industry increasingly pivots toward domain-specific hardware accelerators e.g. tensor processing units (TPUs)~\cite{jouppiInDatacenterPerformanceAnalysis2017}, in-memory compute architectures~\cite{sebastianMemoryDevicesApplications2020, riosInmemoryComputingPhotonic2019}, and optical vector-matrix multipliers~\cite{kalininAnalogOpticalComputer2025}, compatible memory architectures must be researched in tandem.
Optical neural network accelerators have been identified as a particularly promising system, due to their low latency, highly parallel, low-power operation~\cite{mcmahonPhysicsOpticalComputing2023, liangHighclockrateFreespaceOptical2026, xu11TOPSPhotonic2021}.  
However, realizing the full energy advantage of these architectures requires that both data processing \emph{and} data storage remain optical, since repeated conversion to and from electronic memory is power-expensive, limiting the power savings that optical hardware seeks to improve~\cite{aiferSolvingComputeCrisis2025}.
Optical random access memories (ORAMs) are therefore a necessary prerequisite for all-optical computing \cite{alexoudiOpticalRAMIntegrated2020, kazanskiyOpticalComputingStatus2022}. 
\begin{figure}
    \centering
    \includegraphics[width=\linewidth]{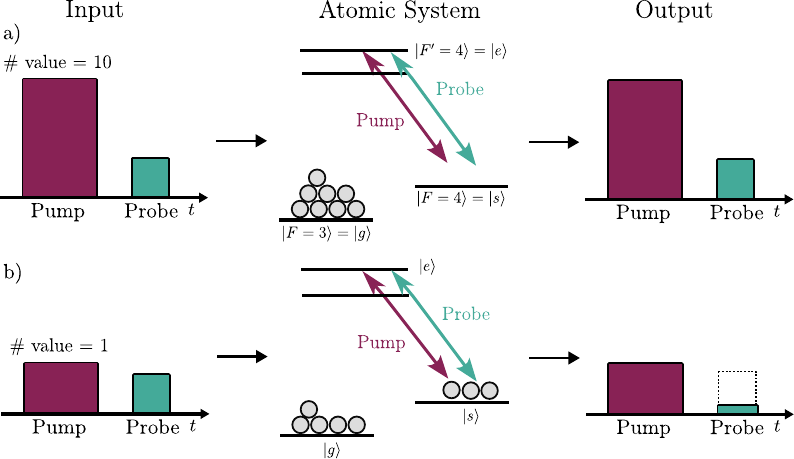}
    \caption{Operating principle of the incoherent atomic ORAM. (a)~A strong pump pulse transfers atoms to the dark state $|g\rangle$, rendering the medium transparent to a subsequent weak probe. (b)~A weaker pump transfers fewer atoms, yielding lower probe transmission. The stored analog value is encoded in the measured probe intensity.}
    \label{fig:pump_probe_principle}
\end{figure}

Within the broad field of optical computing, optical neural networks and specifically optical reservoir computing has been identified as a promising platform for high-throughput, low-energy machine learning \cite{wangOpticalNextGeneration2025}. 
A key characteristic of the reservoir computer (RC) and a distinguishing feature from other architectures, such as an extreme learning machine (ELM)\cite{rausell-campoProgrammablePhotonicExtreme2025}, is its fading memory. Memory may be actively included into optical reservoir computers, explicitly as an element for introducing delay e.g. delay-line systems \cite{abdallaPhotonicReservoirComputing2025}, or alternatively it may arise as an intrinsic feature of the hardware substrate used as a device, often seen in spatially distributed RCs \cite{muletSpatiotemporalModelingOptical2002}. 
Moreover, memories are required to store the information input into reservoir computers. Often, electro-optic devices are used to modulate input information, necessitating storage in the electrical domain; however, all optical architectures seek to remove this conversion. Here, optical memories which can be read or written to optically are required.   
In spatially distributed optical reservoir computers, for example, data are commonly injected via digital micromirror devices (DMDs) or spatial light modulators (SLMs), which are electronically writable and introduce conversion overhead~\cite{dongOpticalReservoirComputing2020, rafayelyanLargeScaleOpticalReservoir2020}. 
Optically addressable SLMs (OASLMs) could address this limitation~\cite{semenovTwocolorOpticallyAddressed2021}, but require a native optical memory to buffer and update weight data. 
In delay-line reservoir computers, the optical fiber loop that provides recurrence acts as a strict first-in first-out (FIFO) queue, fixing virtual node connectivity to a simple ring topology. 
A random-access optical memory replacing the delay-line could enable arbitrary inter-node coupling, enabling investigation into potentially more varied reservoir dynamics. A preliminary investigation into this system is the subject of this work.

Some exemplary physical platforms that have been studied as optically writable ORAMs include phase change materials (PCMs) \cite{gemoPlasmonicallyenhancedAllopticalIntegrated2019, liFastReliableStorage2019, songFabricationfriendlyAllopticalPlasmonicallyenhanced2025}, and different devices that form optical bistable memories \cite{alexoudiOpticalRAMIntegrated2020,ashtianiProgrammablePhotonicLatch2025} for integrated memory. For free-space systems, optically addressable liquid-crystal cells \cite{semenovTwocolorOpticallyAddressed2021, kirzhnerLiquidCrystalHighresolution2014, shresthaHighresolutionOpticallyAddressed2015} and metasurfaces \cite{gongOpticallyAddressedSpatial2021, fanSpatialLightModulator2026} have been investigated. 
Many of these devices have demonstrated successful operation in a research setting, but require further development to be used in real world applications.  

In this work, we employ a room temperature cesium (\Cs) atomic vapor as a free-space storage medium. Compared with fabricated storage platforms, atomic vapor offers high uniformity because all \Cs\ atoms have identical properties, eliminating fabrication-induced variations and making peripheral devices the dominant source of non-uniformity. The inherently continuous three-dimensional spatial structure of the vapor cell also enables scalable multimode operation without the need for microfabrication. Importantly, the nonlinear saturation absorption of the atomic transition provides a native nonlinear response within the memory, enabling in-memory computation and furnishing the activation function for reservoir computing

Here, we present the design, characterization, and reservoir computing application of a free-space \Cs\ vapor ORAM. After introducing the memory operating principle and experimental apparatus (Section~\ref{sec:memory}), we describe the use of a Cs vapor ORAM as an element in a reservoir computing architecture and its performance metrics (Section~\ref{sec:rc}), and evaluate the reservoir against the linear memory capacity, kernel rank, and exclusive-or (XOR) benchmarks (Section~\ref{sec:results}).

\section{Memory Concept and Characterization}
\label{sec:memory}
\subsection{Operating Principle}

The proposed memory encodes a number to be stored in the hyperfine ground-state population distribution of a warm \Cs\ ensemble.
The writing protocol utilizes optical pumping~\cite{happerOpticalPumping1972}, specifically depopulation pumping (or optical shelving), wherein atoms are optically transferred from the upper hyperfine ground state $|s\rangle = |F = 4\rangle$ into a dark state $|g\rangle = |F = 3\rangle$. 
The population pumping results in a polarized ensemble, where the degree population polarization $P$ indicates the magnitude of the number stored.
To read from the memory, the transmission behavior of the probe field is measured as a proxy for the population residing in the dark state. 
The protocol for reading and writing to the memory is illustrated in Figure~\ref{fig:pump_probe_principle}. 
Here, a numerical value is encoded in the atomic vapor using a strong laser pulse tuned to the Cs D1 $F = 4 \rightarrow F^\prime = 4$ transition, with the pulse intensity modulated according to the number to be stored.
To read from the memory cell, a subsequent weak probe pulse of the same frequency but below the saturation intensity ($I_\mathrm{sat}$), is incident on the cell. 
The probe light is transmitted with an efficiency proportional to the population that remains in the $|s\rangle$ state. 
Consequently, higher pump intensities yield greater transmission, representing the storage of larger numbers.
A key feature is the \emph{accumulative} nature of writing in the linear regime: successive pump pulses pump more of the atomic population to the $|g\rangle$ state, provided the pump amplitudes remain below $I_\mathrm{sat}$. 
Furthermore, when the effective pump intensity exceeds the saturation threshold, the ensemble enters a nonlinear regime characterized by its saturation absorption profile.
Memory reset occurs passively as polarized atoms diffuse out of the laser interaction region and are replaced by thermally distributed atoms from the surrounding vapor. While the medium could be actively reset via a repumper laser on the $|F = 3\rangle \rightarrow |F^\prime = 4\rangle$ transition, the necessary calibration and further investigation falls outside the scope of this work.

To store multiple values in a random access manner, we spatially multiplex the Cs vapor cell by deflecting the read and write pulses using an acousto-optic modulator (AOM) \cite{messnerMultiplexedRandomaccessOptical2023}; an example configuration with the two outermost memory rails being addressed simultaneously is shown in the experimental setup (Figure~\ref{fig:experiemntal_reservoir}a). 
By modulating the amplitude of the RF waves driving the AOMs, pulses can be generated. 
A typical modulation waveform for the input AOM (AOM1) and the collection AOM (AOM2) can be seen in Figure~\ref{fig:experiemntal_reservoir}b. 
Different spatial locations can be accessed simultaneously by applying multi-frequency driving waves. 
Modulation waveforms corresponding to the different values to be written to each rail are applied to RF driving frequencies, which determine the spatial location that form individual memory rails. These individual frequency components are summed to access multiple rails simultaneously. 
To avoid confusion, in this work we distinguish between the storage medium (denoted cesium ``cell'') and addressable memory location (memory ``rail'').


\subsection{Experimental Setup}
\begin{figure}[t]
  \centering
  \includegraphics[width=\linewidth]{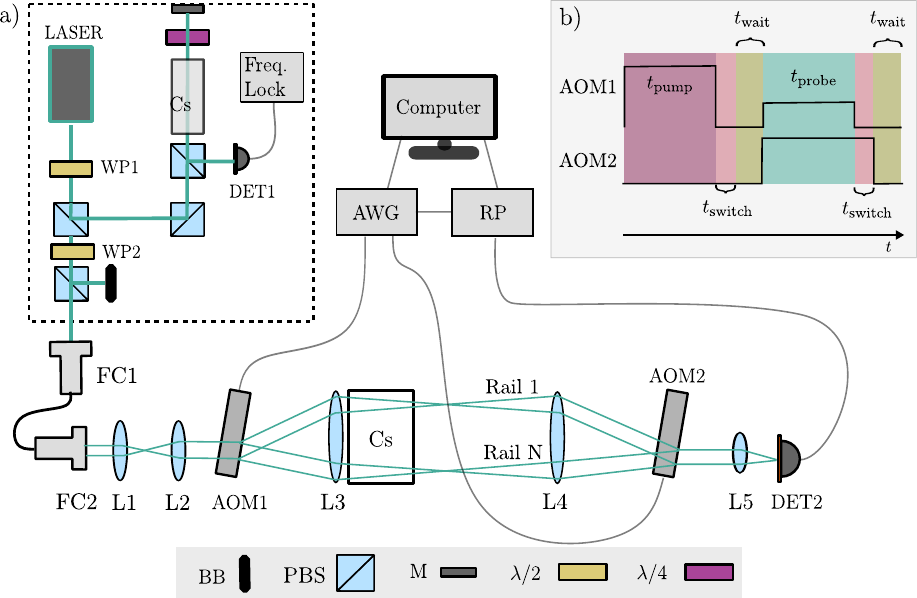}
  \caption{a) Experimental setup of the \Cs\ ORAM. The laser is frequency-locked via Doppler-free spectroscopy (dashed box). After magnification from lenses L1 and L2, the pump/probe beam is diffracted by AOM1 into $N$ spatial rails that traverse the Cs cell; AOM2 recombines the modes onto a single detector path. FC: Fiber coupler; L: Lens; DET: Detector; AWG: arbitrary wave generator; PBS: polarizing beamsplitter; BB: beam block; M: mirror; RP: Red Pitaya; WP: waveplate; $\lambda/2$: half-wave plate; $\lambda/4$: quarter-wave plate. b)~Single-rail pump-probe timing sequence. AOM2 is turned off during pumping to prevent damage to the detector. }
  \label{fig:experiemntal_reservoir}
\end{figure}

The full experimental setup is shown in Fig.~\ref{fig:experiemntal_reservoir}. 
An external-cavity diode laser (ECDL; Sacher Lasertechnik Micron) is locked to the $F=4\!\to\!F'=4$ Lamb-dip of a Doppler-free spectroscopy using a Red Pitaya STEMlab 125-14 running the open-source \textit{Linien} software~\cite{wiegandLinienVersatileUserfriendly2022}. 
The beam is then expanded by a telescope formed of L1 and L2 to a $1/e^2$ diameter of $\SI{3.1}{\milli\metre}$ and directed to the first acousto-optic modulator (AOM1; AA Optoelectronics DTSX-400-900). 
The beam power is adjusted to $P_\mathrm{in} = 10\, \mathrm{mW}$ at the fiber coupler output (FC2, Schäfter + Krichoff 60FC-SP-4-M8-10) using a polarizing beamsplitter (PBS) and a half-wave plate (WP2).
AOM1 is driven by a multi-tone RF signal from a Zurich Instruments HDAWG-4 arbitrary waveform generator, which simultaneously generates eight carrier frequencies ($f_\mathrm{RF} = 75, 78, \ldots, \SI{96}{\mega\hertz}$, separated by $\SI{3}{\mega\hertz}$ to avoid crosstalk), each amplitude-modulated independently to produce pump and probe pulse pairs at eight distinct rails within the vapor cell. 
A 4f relay system (lenses L3, L4; $f=\SI{1}{\metre}$) images the spatially separated beams through the Cs cell (Precision Glassblowing; $\SI{10}{\centi\metre}$ length, $\SI{5}{\centi\metre}$ diameter, $\SI{10}{\torr}$ N$_2$ buffer gas at $\SI{30}{\degree}$). 
A second AOM (AOM2) recombines all spatial modes onto a single avalanche photodetector (DET2; Thorlabs APD430A/M) and acts as an optical gate to protect the detector during high-power pump pulses (Fig.~\ref{fig:experiemntal_reservoir}b). 
Probe traces are acquired using the deep acquisition mode of a Red Pitaya STEMlab 124-14 board and transferred to a host computer for further application specific processing.

In this work, unless otherwise stated, the rails are pumped for $t_\mathrm{pump } = 10\, \mu$s, and probed for $t_\mathrm{probe} = 10\, \mu$s, with switching and waiting intervals of $t_\mathrm{switch} = 2\, \mu$s and $t_\mathrm{wait} = 3\, \mu$s. 
This yields a memory operation time of $t_\mathrm{pp} = 30\, \mu$s, defined as the time required to execute a single read and write operation.
Notably, the architecture supports simultaneous write operations across multiple rails.
The amplitude of the probe pulses, corresponding to the value read out of our memory is determined by averaging over a $2\, \mu$s subsection centered with the probe interval of the measured pulse.  
In this system, we deflect the beams in a single axis, with driving electronics limiting the number of resolvable spots to 8 rails.
Substituting single axis AOMs for 2D AOMs and corresponding driving electronics would enable access to a total of 64 memory modes such that the cell aperture becomes the limiting factor. 

\subsection{Memory Characterization}
\begin{figure}[t]
    \centering
    \includegraphics[width=\linewidth]{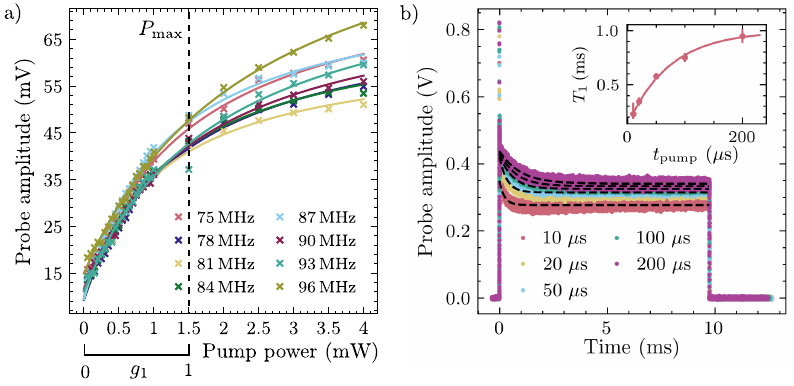}
    \caption{Characteristic memory responses (a) Saturation absorption transfer function of the eight ORAM rails. Experimental data are fit with a saturable absorber profile, which defines the nonlinear function $\theta$ used in reservoir simulations. Mean saturation power: $P_\mathrm{sat} = 1.5~$mW.  (b) Decay of atomic polarization for pump durations $t_\mathrm{pump} \in \{10, 20, 50, 100, 200\}~\mu$s at $P_\mathrm{pump} = 1.5~$mW. The peak feature at $t=0$ is leaked pump light indicating the switch from the pump to the probe pulse. \textbf{Inset:} Measured $T_1$ vs.\ $t_\mathrm{pump}$, fit with $T_1 = a(1 - e^{-kx}) + b$; the curve saturates at approximately $1~$ms for $t_\mathrm{pump}  \geq 200\, \mu$s.}
    \label{fig:charatersitics}
\end{figure}

To characterize the memory operation, the pump pulse energy is varied while monitoring the response of the probe pulse. 
The experimental sequence for each rail consisted of an initial $10\, \mathrm{\mu s}$ probe pulse, a $5\, \mathrm{\mu s}$ delay, a $10\, \mathrm{\mu s}$ pump pulse, and a second $5\ \mathrm{\mu s}$ delay before a second $10\ \mathrm{\mu s}$ probe, for each rail. 
The transmitted intensity of this second probe pulse is shown in Figure~\ref{fig:charatersitics}. 
To ensure consistency, the AOM2 RF driving power is normalized such that a single probe pulse yielded a uniform photodiode voltage. 
Throughout the measurements, the probe power is maintained at $50\, \mathrm{\mu W}$, and the pump power is varied between $0-1\,\mathrm{mW}$ in $50\, \mathrm{\mu W}$ increments, and scanned from $1.5 - 4\, \mathrm{mW}$ in $500\ \mu W$ steps. 
This increased step size at higher powers is adopted to optimize data acquisition time.
The response of the probe transmission to increasing pump power follows a saturable absorber profile of the form $\theta(P_\mathrm{pump}) = a\frac{P_\mathrm{pump}}{1+P_\mathrm{pump}/P_\mathrm{sat}} + b$. 
In the linear regime ($P_\mathrm{pump} \ll P_\mathrm{sat}$), probe transmission increases approximately linearly with pump energy. 
Successive pump pulses result in the accumulation of the population, resulting in in-memory addition, provided the pulses are weak enough to not enter the saturated regime.  
At higher pump powers, the atomic population available for transfer to $|g\rangle$ saturates, producing a rectified response.
We measure the average pump saturation power across all rails as $P_\text{sat} = 1.5\, \mathrm{mW}$.
We see that the normalization of the rails at the edges of the AOM operating range, and for the $\SI{87}{\mega\hertz}$ rail was not successful. 
We attribute this to large variation in the diffraction power of the outermost rails. 
The maximum pump power is set to $P_\mathrm{max} = P_\mathrm{sat}$ to access the onset of saturation while remaining within a bounded dynamic range. A tunable input scaling factor $g_1 \in [0,2.4]$ controls the fraction of $P_\mathrm{max}$ applied to each pulse, allowing systematic variation of the effective nonlinearity.

To quantify the resolution of the system, we measure the difference in transmitted probe amplitude between the pumped and unpumped states, for a pump pulse of $P_\mathrm{pump}= 1.5\, \mathrm{mW}$. 
We determined that a probe power of $P_\mathrm{probe} = 50\, \mu \mathrm{W}$ provides the largest dynamic range for a cell temperature $T = 30^\circ\, \mathrm{C}$, which is the probe power used in the rest of this work unless otherwise specified. 
Using a pump pulse duration $t_\mathrm{pump} = 10\ \mu \mathrm{s}$, we measured a difference in detected photodiode voltage  of between $U = 0.15\, \mathrm{mV}$ and $U = 0.44\ \mathrm{mV}$, with a standard deviation of $\sigma = 1\, \mathrm{mV}$. 
The observed fluctuations arise from technical noise in the AOMs, where thermal variations in deflection efficiency are most pronounced at low RF driving powers. 
These effects may be further reduced through temperature stabilization of the AOMs or by decoupling the pulse generation process and the deflection to different rails.
This yields $N = 14.5$ resolvable levels, or an effective resolution of $\log_2 N = 3.8$ bits. 
This resolution is time-dependent; the value reported here corresponds to a readout performed $3 \mathrm{\mu s}$ after pumping; we discuss the decoherence mechanism further in the next section.

\subsection{Memory Lifetime and Time-Bandwidth Product}\label{sec:lifetime}

The stored population decays as polarized atoms diffuse out of the laser interaction volume and are replaced by atoms in thermal equilibrium. 
This effect occurs continuously from the moment the atoms are pumped, thus meaning the polarization state of the atoms is continuously depleting. 
Charge depletion is a known phenomenon in electronic dynamic random access memory (DRAM), overcome by periodic re-charging. 
The diffusion-limited lifetime is estimated as $\tau_\mathrm{mem} = d_\mathrm{beam}^2 / 4D \approx 1.3~$ms for our beam diameter $d_\mathrm{beam} = 3.1~$mm and Cs-in-N$_2$ diffusion coefficient $D = \SI{17(10)}{\centi\metre\squared\per\second}$. 
We measure the $1/e$ lifetime $T_1$ by pumping a single rail (the central rail at $81~$MHz) with for $10\, \mu$s, with a pump power of $\SI{1.5}{\milli\watt}$. The transmission of a $\sim\!10~$ms weak probe pulse ($P_\mathrm{probe} = 9\, \mu$W) is measured for pump pulses of varying durations.

The measured $1/e$ lifetimes are $T_1 = \{230(50),\, 340(40),\, 570(30),\, 750(40),\, 950(60)\}~\mu$s for pump durations $t_\mathrm{pump} = \{10, 20, 50, 100, 200\}~\mu$s respectively.  We observe that the duration of the pump saturates to approximately $\SI{1}{\milli\second}$ with increasing pump pulse duration, as show in the inset of Fig.~\ref{fig:charatersitics}). 

To quantify the information storage capacity, we define the memory time-bandwidth product $\mathrm{TBP}_\mathrm{mem}$ as the number of pump-probe operations executable within the $1/e$ lifetime:
\begin{equation}
  \mathrm{TBP}_\mathrm{mem} = \frac{T_1}{t_\mathrm{pp}},
\end{equation}
where $t_\mathrm{pp}$ is the duration of a single pump-probe cycle. For our system, $t_\mathrm{pp} = 30~\mu$s (including a $10~\mu$s pump, $10~\mu$s probe, and switching/wait times), yielding $\mathrm{TBP}_\mathrm{mem} = 7.7$ at $T_1 = 230~\mu$s. 
This presents a second limitation in the number of memory rails which are accessible within the storage time. 
To increase the number of accessible memory modes, one must either increase the storage time or decrease the time of a single read-write (pump-probe) operation.
The storage time may be extended using an array of miniaturized Cs cells, where the pump/probe beam fully encompasses the atomic volume such that the atoms cannot diffuse out of the interaction region. 
These cells must be filled with an anti-relaxation coating to prevent collisions with the cell walls becoming a limiting mechanism and are a topic of current active research \cite{mottolaOpticalMemoryMicrofabricated2023, nodopAtomicVaporCells2026, liWaferscaleFabricationTemperaturecompensated2025}.
The duration of the pump and probe sequence could be reduced by utilizing faster AOMs with rise times on the order of tens of nanoseconds, thereby enabling shorter pulses. However, this improvement would come at the expense of decreasing the total number of addressable rails.

\begin{table}[ht]
    \centering
    \begin{tabular}{|l|l|}
    \hline
    Figure of merit & Value  \\
    \hline
     Memory lifetime ($T_1$) & $230\, \mathrm{\mu s}$ \\
     Resolution & 3.8 bits \\
     Number of rails ($N$) & 8 \\ 
     Read/write duration ($t_\mathrm{pp}$) & $30\, \mathrm{\mu s }$ \\
     Memory bandwidth ($\mathrm{TBP}_\mathrm{mem}$) & 7.7\\
    \hline
    \end{tabular}
    \caption{The key characteristics of the incoherent memory experiment.}
    \label{tab:Incoherent_figures_of_merit}
\end{table}


The key figures of merit are summarized in Table~\ref{tab:Incoherent_figures_of_merit}. Having established the characteristics of the medium as a memory element, we now describe the reservoir computing architecture implemented in this work.
\section{Reservoir Computing Architecture and Metrics}
\label{sec:rc}

\subsection{Memory-Based Reservoir Computing}

Physical reservoir computing (RC) exploits the nonlinear transient dynamics of a fixed physical system to map low-dimensional inputs into a high-dimensional state space, after which only a single linear readout layer requires training~\cite{jaegerEchoStateApproach2001}. 
Optical delay-line reservoirs achieve recurrence via a fiber loop that functions as a FIFO queue~\cite{largerPhotonicInformationProcessing2012}; virtual nodes are accessed sequentially, and connectivity is topologically fixed. 
By replacing the delay line with a multi-rail ORAM, arbitrary read/write scheduling becomes possible, enabling flexible intra-node coupling.
In temporally multiplexed systems such as delay-line reservoirs and the system presented in this work, information is injected into and read sequentially from the reservoir. 
Hence, we define the time scales relevant to the system in question; the input dataset $\mathbf{u}$ is injected into the reservoir as a piecewise function, with each element of $\mathbf{u}$ being injected with duration $T$, indexed by a discrete time $t$. 
Consequently, $\mathbf{u}(t=0)$ corresponds to injection of the first logical data point. 
The number of data points injected into the reservoir is given by the training (test) set size $N_\mathrm{train} (N_\mathrm{test})$.
We define virtual nodes by further subdividing the clock cycle into $N_v$ distinct segments, where $N_v$ is the length of a mask weight vector $\mathbf{w}_\mathrm{mask}$. 
Each masked element is injected into the reservoir with duration $T_v$, such that $N_v = T/T_{\mathrm{v}}$, as is typical in temporally multiplexed reservoir computing \cite{appeltantInformationProcessingUsing2011, hulserRoleDelaytimesDelaybased2022, ortin5TimeDelay2019}.
Experimentally, we set $T_v = t_\mathrm{pp} = 30\, \mu$s, the time to execute a single read and write operation on a memory rail. 
The value of a single element injected into the reservoir is given by: $\mathbf{j}(k) = \mathbf{u}(t) \cdot \mathbf{w}_\mathrm{mask} (k^\prime) $ where $k^\prime \in (0,N_v] $ is the mask or virtual node index, and the system index $k = N_vt + k^\prime$.
The mask is drawn from a uniform distribution in the interval $[0,1]$; differing from the convention for the mask interval of [-1,1] as our hardware cannot represent negative numbers. 
The input layer of the reservoir computer is depicted in Figure~\ref{fig:2_memory_res_scheme}

We inject the masked input sequentially into the reservoir and fill the state matrix $\mathbf{S}$ with responses. The state matrix is then normalized such that the entries are between $\mathbf{x}_\mathrm{out}  \in [0,1]$.
We include an additional bias vector, resulting in a $N_\mathrm{train} \times (N_v + 1)$  dimensional state matrix. 
This is shown in Figure~\ref{fig:2_memory_res_scheme} as the output layer.
To train the reservoir an output weight vector $\mathbf{w}_\mathrm{out}$ is found that minimizes the cost function: 
\begin{equation}
    \arg\min_{\mathbf{w}_{\mathrm{out}}} \left(||\mathbf{S}\mathbf{w}_{\mathrm{out}} + \mathbf{y} ||_2^{2}  - \lambda_{\mathrm{reg}}||\mathbf{w}_{\mathrm{out}}||_2^2 \right)
\end{equation}\label{eq:opt_prob}
where $||\cdot ||_{2}$ is the Euclidean norm, $\lambda_\mathrm{reg}$ is the regularization parameter and $\mathbf{y}$ is the true value of the task for inputs $\mathbf{u}$. 

Having discussed information injection and population of the state matrix from the reservoir output, the next section details how to utilize the memory presented in Section~\ref{sec:memory} as a reservoir. The following sections then describe the training process and the metrics used to evaluate the reservoir's performance.

\subsection{Reservoir Model}

\begin{figure}[t]
  \centering
  \includegraphics[width=\linewidth]{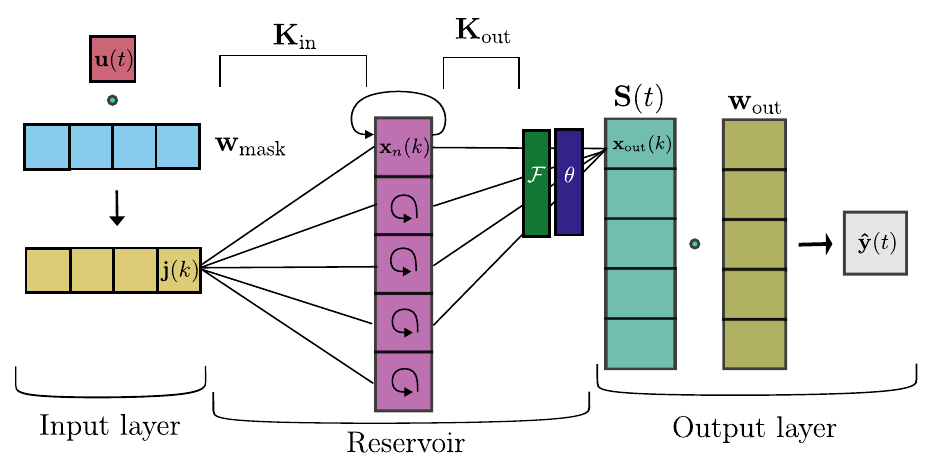}
  \caption{Schematic of the memory-based reservoir computer. At each time step, the memory rails are first probed (read) before being pumped (written). The coupling matrices $\mathbf{K}_\mathrm{in}$ and $\mathbf{K}_\mathrm{out}$ determine which rails are addressed, establishing the effective network topology.}
  \label{fig:2_memory_res_scheme}
\end{figure}

In our reservoir architecture, the physical memory rails are temporally multiplexed to create a high-dimensional output state space. 
The masked input $\mathbf{j}(k)$ is written to a rail determined by the input coupling matrix $\mathbf{K}_\mathrm{in}$.  
$\mathbf{K}_\mathrm{in}$ contains values between $[0,1]$, which represent the strength with which a specific memory rail should be written to at timestep $m$, where $m = k \mod M$. $M$ is a coupling matrix repetition index, defining after how many system time steps the coupling matrix should repeat, which we set to $M = N = 8$.
A similar approach using coupling matrices to determine the input and output coupling was presented in Ref.~\cite{jaurigueReservoirComputingDelayed2021}. We model the state of a specific rail $x_n$, as the effective pump energy of that rail; an overview of the architecture can be seen in Figure~\ref{fig:2_memory_res_scheme}.
As the atomic population state decays with the storage time, the state of the memory rail from the previous timestep is multiplied by a decay factor $\mathcal{F}(\Delta k_n)$, modeling the exponential population decay between successive writes. This is given by the exponential decay as observed in Section~\ref{sec:lifetime}, where $\mathcal{F}(\Delta k_n) = \exp(-\Delta k_n\, t_\mathrm{pp}/T_1)$ and $\Delta k_n$ is the number of steps elapsed since rail $n$ was last written.

The state of rail $n$ evolves as:
\begin{equation}
  \mathbf{x}_n(k{+}1) = \bigl(\mathbf{K}_\mathrm{in}\bigr)_n^m \!\left(g_1 \mathbf{j}(k{+}1)P_\mathrm{max}\right) + \mathcal{F}(\Delta k_n)\cdot\mathbf{x}_n(k),
\end{equation}
At each timestep, we can also choose the memory rails which are read from by defining the $\mathbf{K}_\mathrm{out}$ coupling matrix. 
The reservoir output is:
\begin{equation}
  \mathbf{x}_\mathrm{out}(k) = \sum_{n=1}^{N} \theta\!\left(\bigl(\mathbf{K}_\mathrm{out}\bigr)_n^m \cdot \mathcal{F}(\Delta k)\cdot\mathbf{x}_n(k)\right),
\end{equation}
where $\theta$ is the empirically measured characteristic saturation absorption transfer function (Fig.~\ref{fig:charatersitics}). 
The states $\mathbf{x}_\mathrm{out}$ are then used to populate the state matrix $\mathbf{S}$. 
Note that the memory cells are first read from before they are written to. 
Consequently, we set $\mathbf{K}_\mathrm{out}$ to the identity matrix, with dimension $N$, rolled over by one, such that the final memory cell is read first. 
This results in reading the cell that was just written out at each time step, increasing the effective memory capacity of the system. 
. 

In this work, the network topology is controlled by varying ($N_v$) relative to ($N$). A mismatch such that ($N_v = N \pm 1$) shifts the rail-to-virtual-node mapping at each clock cycle, thereby producing a ring-coupling structure between successive virtual nodes. 
In this preliminary investigation, only the number of virtual nodes is varied to modify the reservoir topology, while $\mathbf{K}_\mathrm{in}$ and $\mathbf{K}_\mathrm{out}$ are kept fixed to limit the scope of the study.
Aside from the coupling, we may change the reservoir operations behavior by varying the linearity of the response of a single write action by changing the coefficient $g_1$. 
Furthermore, by changing the duration between the pump and probes $t_\mathrm{wait}$, we can vary the memory capacity of the system. 
It is useful to define the logical memory capacity $T_\mathrm{logic}$, which corresponds to how many logical inputs $\mathbf{u}$ can be input into the reservoir before the state of the first memory cell written to decays to its $1/e$ value. 
Thus we define:
\begin{equation}\label{eq:tbp}
     \mathrm{TBP}_\mathrm{logic}  =\frac{T_1}{N_v T_v} =\frac{ \text{TBP}_\text{mem}}{N_v}. 
\end{equation}
The minimum number of virtual nodes considered is $N_v = 3$, yielding a maximum $\mathrm{TBP}_\mathrm{logic} = 2.6$.
These variables, $N_v$, $g_1$ and $T_\mathrm{mem}$, define the performance parameter space we explore in this work. 

\subsection{Training}
The output weight vector $\mathbf{w}_\mathrm{out}$ is trained using ridge regression on a state matrix $\mathbf{S}$ assembled from $N_\mathrm{train} \approx 300$ input samples. 
Performance is evaluated on an unseen test set of $N_\mathrm{test} = 300$ samples. 
The size of the train and test set varies by approximately $10\%$ due to data set segmenting; a single experimental run of the reservoir required multiple experimental traces. We find this to have no appreciable effect on system performance.  

To train the reservoir, we solve for $\mathbf{w}_\mathrm{out}$ using ridge regression on the state matrix generated from the train set. 
This yields a closed-form solution to the minimization problem given by Eq. \ref{eq:opt_prob}: 
\begin{equation}
    \mathbf{w}_{\mathrm{out}} =( \mathbf{S}^T \mathbf{S} - \lambda_{\mathrm{reg}} \indicator )^{-1} \mathbf{S}^T \mathbf{y},  
\end{equation}
where $\mathbf{S}^T$ is the transpose of the state matrix, $\indicator$ is the identity matrix and $\lambda_{\mathrm{reg}}$ is the regularization parameter, set to $\lambda_{\mathrm{reg}} = 5 \times 10^{-6}$.

\subsection{Performance Metrics}
The reservoir's performance is measured by its linear memory capacity (LMC), kernel rank (KR), and its ability to predict the XOR between current and previous inputs. The reported values represent an average over five different $\mathbf{w}_\mathrm{mask}$ configurations, where the reported error bars indicate the standard deviation. We discuss these metrics in detail below, beginning with the LMC.

\paragraph{Linear Memory Capacity (LMC)} 
The LMC quantifies how well the reservoir is able to recall past inputs \cite{wringeReservoirComputingBenchmarks2025}. 
To measure the LMC, we inject a random string of numbers $\mathbf{u} $ drawn from a uniform distribution $U[0,1]$ into the reservoir. 
A reservoir is trained to recall the input $\mathbf{u}$ $n$ time steps in the past $\mathbf{y}(t) = \mathbf{u}(t-n)$ and the performance of the reservoir in recalling that input is measured by the capacity $C_n$:
\begin{equation}
   C_n = 1 - \mathrm{NMSE}(\mathbf{y}_n,\mathbf{\hat{y}}_n) =  1- \frac{\sum_{k=1}^{N_\mathrm{test}} \| \hat{y}^k_n- y^k_n \|^{2}}{N_\mathrm{test} \; \mathrm{Var} (\mathbf{y}_n)},
\end{equation}
where  $\mathbf{y}_n$ are the predictions of the input $n$ steps ago and $\mathrm{Var}(\mathbf{y_n})$ is the variance of the target data set. 
The linear memory capacity is then the sum of all the time delays considered.
\begin{equation}
    \text{MC}  = \sum_{n =1 }^{N_\mathrm{lim}} C_{n}.
\end{equation}
We choose the maximum evaluation time $N_{\mathrm{lim}} = 5$. 
This is sufficient as it exceeds the maximum time-bandwidth product ($\mathrm{TBP}_\mathrm{logic} = 2.6$) of the system. 

\paragraph{Kernel Rank (KR)} 
The KR quantifies the dimension of the feature space into which the reservoir projects the input information \cite{wringeReservoirComputingBenchmarks2025}. To measure KR, a sequence of random inputs $\mathbf{u}$ is randomly drawn from a uniform distribution $U [0,1]$ and injected into the reservoir. 
The resulting reservoir states are collected into a state matrix $\mathbf{S}$, and by performing singular value decomposition, the kernel rank can be determined by taking the rank of the eigenvalue matrix ~$\mathbf{\Sigma}$:
\begin{equation}
    \mathbf{S} = \mathbf{U} \mathbf{\Sigma} \mathbf{V}^{\dagger},
\end{equation}
\begin{equation}
    r_\mathbf{S} = \operatorname{rank} (\mathbf{\Sigma}),
\end{equation}
where $\mathbf{U}$ and $\mathbf{V}$ are unitary matrices containing the left and right singular vectors of the state matrix. 
In practice, due to noise and experimental imperfections, singular values are rarely exactly zero. 
Therefore, when reporting KR, we provide a threshold below which SVD values are considered zero set, in this case to 2\% of the maximum singular value. 
The theoretical upper bound of the kernel rank is determined by the number of virtual nodes in the system, here varied between $N_v \in [8,14]$.

\paragraph{XOR Bit Error Rate (BER)} 
The XOR logic gate is a standard benchmark for reservoir computing systems because its truth table is nonlinearly separable. 
Consequently, a linear classifier cannot solve this task without the reservoir successfully mapping the input into a higher-dimensional feature space.
The task consists of computing the XOR function of two consecutive inputs binary inputs, the target function is thus:
\begin{equation}
    \mathbf{y}(t) = \mathrm{XOR} (\mathbf{u}(t), \mathbf{u}(t-1)),
\end{equation}
where the input sequence consists of binary values $\mathbf{u}\  \in \{0,1\}$. 
The prediction performance is quantified by the bit error rate (BER), defined as the ratio of false classifications to the total number of samples in the test set: 
\begin{equation}
\text{BER} = \frac{N_\text{wrong}}{N_\text{test}},
\end{equation}
where $N_\text{wrong}$ is the number of misclassified inputs and $N_\text{test}$ is the size of the test set.
Since the reservoir prediction $\mathbf{\hat{y}}(t)$ is a continuous variable, it is discretized into binary classes by applying a threshold.
Specifically, predictions $\geq 0.5$ are assigned to 1 and $< 0.5 $ are assigned to $0$.

\section{Results and Discussion}
\label{sec:results}

We evaluate the reservoir performance over three independent parameter axes: network topology ($N_v$), the decay timescale ($\mathrm{TBP}_\mathrm{mem} = T_1/t_\mathrm{pp}$, varied via $t_\mathrm{wait}$), and the input nonlinearity ($g_1$). 

\subsection{Linear Memory Capacity}\label{sec:LMC}

\begin{figure}[t]
  \centering
  \includegraphics[width=\linewidth]{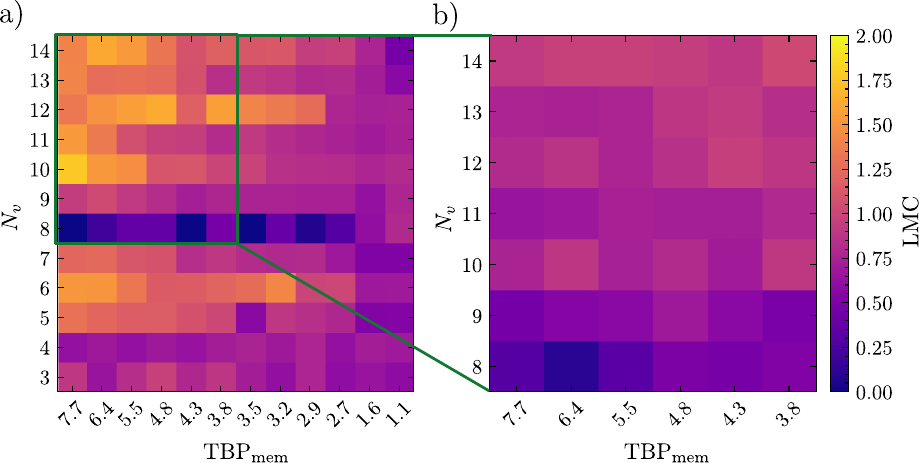}
  \caption{Linear memory capacity as a function of virtual node number $N_v$ and $\mathrm{TBP}_\mathrm{mem}$, for $T_1 = 230~\mu$s ($t_\mathrm{pump}=10~\mu$s), $g_1 = 1$. (a)~Simulation; (b)~experiment. The experimentally scanned region is indicated in green. The maximum experimental MC$= 1.0\pm0.2$ is achieved at $N_v = 14$, $\mathrm{TBP}_\mathrm{mem} = 3.8$.}
  \label{fig:2_LMC_nv_dt}
\end{figure}
To understand the memory behavior of the reservoir, we first consider the linear memory capacity as a function of $N_v$ and $\mathrm{TBP}_\mathrm{mem}$, as shown in Figure~\ref{fig:2_LMC_nv_dt}. 
From simulation (Figure~\ref{fig:2_LMC_nv_dt}a), we predict the highest achievable memory capacity to be $\text{LMC} = 1.8(5)$ at $t_\mathrm{wait} = 3\, \mu \mathrm{s}\ (\mathrm{TBP_\mathrm{mem} = 7.7})$, $N_v = 10$ corresponding to a $\mathrm{TBP}_\mathrm{logic}$ of $0.8$.
Experimentally, we find that the optimal operating point is $ N_v = 14,\ t_\mathrm{wait} = 18 \, \mu\mathrm{s}\ (\mathrm{TBP_\mathrm{mem} = 3.8})$ yielding $\text{LMC} = 1.0(2)$, which corresponds to a TBP of $\mathrm{TBP}_\mathrm{logic} = 0.29$.
This indicates that the time bandwidth product alone is not the sole indicator for memory capacity. A sufficiently high dimensionality in the state space is required to separate the trajectories corresponding to different time scales; the dimension of the network ($N_v$) influences the linear memory capacity. 
Consequently, lower dimensional systems ($N_v<N$) perform poorly despite having a higher $\text{TBP}_\text{logic}$. Both simulation and experiment show improved memory capacity for virtual node numbers of $N_v = 10, 12, 14$.
We also observe a performance reduction due to resonances within the reservoir; when the number of virtual nodes is an integer number, or a simple fraction of the number of memory rails,  $N_v = N = 8$, or  $N_v = 0.5N$, the reservoir performance drops significantly in both simulation and experiment. 
These resonance regimes are known to degrade performance in delay-based systems, due to a reduction in the effective richness in the reservoir dynamics \cite{kosterLimitationsRecallCapabilities2020, hulserRoleDelaytimesDelaybased2022, rohmReservoirComputingUsing2020}. 

The fundamental constraint is that the logical time-bandwidth product for $N_v > 8$ at $\mathrm{TBP}_\mathrm{logic} < 1$ under current operating conditions. This indicates that the state of the first virtual node decays to below $1/e$ before the complete set of virtual nodes comprising a single logical input has been injected—severely restricting the temporal depth of the reservoir. For comparison, delay-line reservoirs routinely achieve LMCs of order 10 or more~\cite{kosterLimitationsRecallCapabilities2020}.
Investigations into extending the memory capacity by increasing the pump time to $t_\mathrm{pump} = 200~\mu$s ($T_1 = 950~\mu$s) yielded no improvement; the maximum drops to LMC$= 0.36\pm0.18$. 
Although a longer lifetime increases $\mathrm{TBP}_\mathrm{mem}$, the accompanying increase in $t_\mathrm{pp}$ reduces $\mathrm{TBP}_\mathrm{logic}$ (Eq.~\ref{eq:tbp}), negating the benefit. 
Thus, to achieve higher LMC, we must find alternate means to increase the $\mathrm{TBP}_\mathrm{mem}$ of the system, for example by reducing $t_\mathrm{pp}$ by using AOMs with a shorter rise time and with higher energy pump pulses.   
The limited LMC is characteristic of an extreme learning machine (ELM) operating regime rather than a true recurrent reservoir.

\subsection{Kernel Rank}

\begin{figure}[t]
  \centering
  \includegraphics[width=\linewidth]{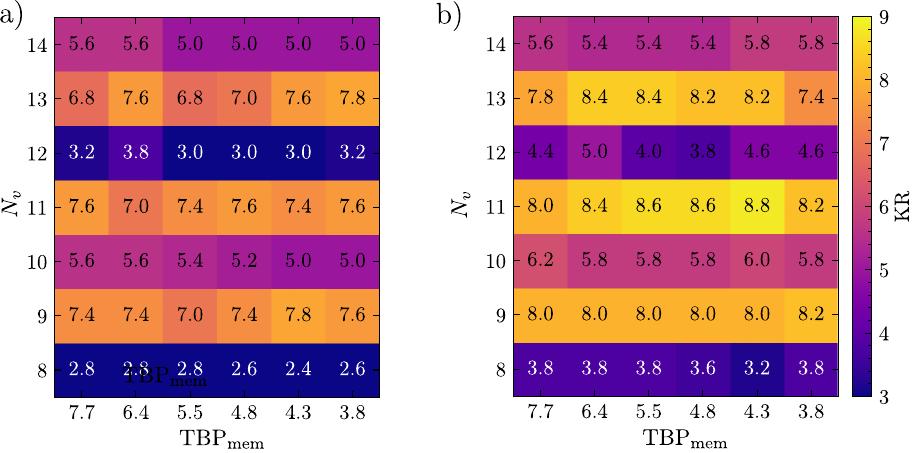}
  \caption{Kernel rank as a function of $N_v$ and $\mathrm{TBP}_\mathrm{mem}$ for $T_1 = 230~\mu$s. (a)~Simulation, KR$_\mathrm{sim}^\mathrm{max} = 7.8\pm0.4$; (b)~experiment, KR$_\mathrm{exp}^\mathrm{max} = 8.8\pm0.4$.}
  \label{fig:2_KR_nv_dt}
\end{figure}

The dependence of the kernel rank on the reservoir parameters $N_v$ and $\text{TBP}_\text{mem}$, for a short memory lifetime $T_1 = 230\,\mathrm{\mu s}$ is shown in Figure \ref{fig:2_KR_nv_dt}.
We observe the KR is primarily determined by the number of virtual nodes, reducing the memory TBP of the system has a negligible effect on the dimensionality of the projection in both simulation and experiment. Experimentally, a maximum of $\mathrm{KR}= 8.8\pm0.4$ is reached, approaching the theoretical upper bound of $N_v$.
However, we do not anticipate the higher experimental KR to be indicative of higher-dimensionality performance compared to the simulated reservoir; rather, we expect that further noise correlations exist, which indicate that a threshold of 2\% may be too small and thus overestimate the experimental kernel rank.
Further investigation similar to the analysis presented in Ref.~\cite{skalliComputationalMetricsParameters2022} is required to confirm this but was deemed beyond the scope of this work.

\begin{figure}[t]
  \centering
  \includegraphics[width=0.6\linewidth]{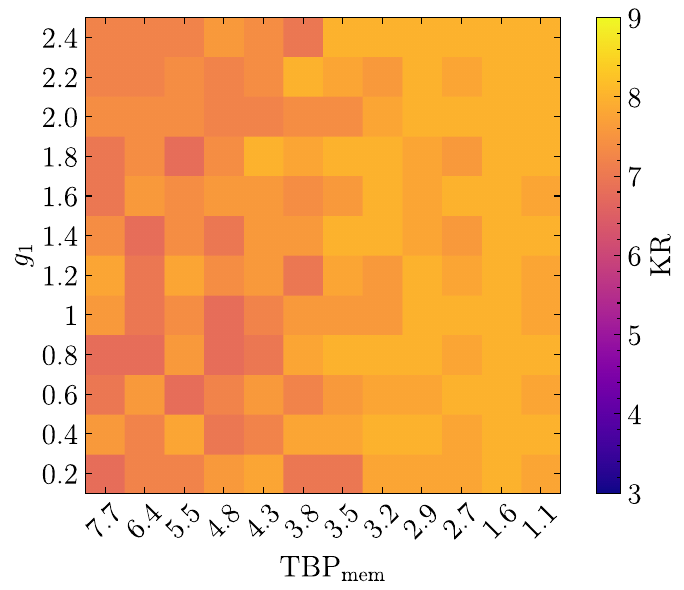}
  \caption{Simulated kernel rank as a function of input scaling $g_1$ and $\mathrm{TBP}_\mathrm{mem}$ for $N_v = 11$, $T_1 = 230~\mu$s. $g_1 < 1$ restricts pumping to the linear regime. The KR range spans $6.8$--$8.0$, with no strong dependence on $g_1$.}
  \label{fig:2_KR_g1}
\end{figure}

We investigate the influence that varying the operation regime between linear and nonlinear systems has on the reservoir by varying $g_1$, revealing that varying the pump power—and thus the degree of saturation nonlinearity—has a weak effect on KR. The results can be seen in Figure~\ref{fig:2_KR_g1}, where we observe the KR values ranging only between $6.8$ and $8.0$ across the full range of $g_1$. This indicates that the dimensional expansion of the reservoir state is driven primarily by the coupling topology (the choice of $N_v$ relative to $N$), rather than by the atomic saturation nonlinearity.
These results suggest that the saturation absorption profile provides insufficient nonlinear transformation for robust dimensionality expansion independent of topology. An investigation into whether the incorporation of an additional nonlinear element—such as a semiconductor optical amplifier—could further increase the nonlinearity remains an extension to this work.

\subsection{XOR Classification}

\begin{figure}[b]
    \centering
    \includegraphics[width=\linewidth]{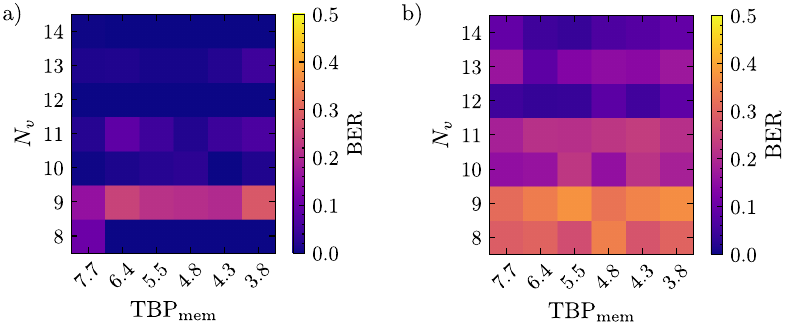}
    \caption{(a) Simulated and (b) experimental XOR performance (BER) as a function of $N_v$ and $\mathrm{TBP}_\mathrm{mem}$ for $T_1 = 230~\mu$s, $g_1 = 1$.
    In simulation $\mathrm{BER_{\mathrm{sim}}^{\text{(min)}}} = 0$, for $N_v \in \{8, 12, 14\}$, for all values of  $\text{TBP}_\text{mem}$, except $\text{TBP}_\text{mem} = 7.7$. 
   The lowest experimental accuracy is $\mathrm{BER}_{\mathrm{exp}}^\text{(min)}= 0.03(3)$ is obtained at $N_v = 12$, $\text{TBP}_\text{mem}= 6.4$.}
    \label{fig:2XOR_nv_dt}
\end{figure}

We first investigate the effect of $N_v$ and $\mathrm{TBP}_\mathrm{mem}$ on the performance of XOR task, as can be seen in Figure~\ref{fig:2XOR_nv_dt} .
The test size of $300$ samples imposes a quantization floor on the error metric; the minimum resolvable BER is $1/300 \approx 0.003$. 
Extending the size of the data set was not feasible in this work due to the experimental limitations from device data transfer. 
Importantly, this means that a $\text{BER} = 0$, as we achieve in the simulation, requires reevaluation with longer test sets to get an accurate measure of the performance.
Optimal performance is obtained for even values of $N_v$, consistent with the correlation between even-$N_v$ topologies and higher LMC observed in Section~\ref{sec:LMC}. 
This confirms that memory depth, rather than projection dimensionality, is the limiting factor for the XOR task, which requires an output representation containing the previous time step. 
The best performing configuration while scanning $N_v$ and $\mathrm{TBP}_\mathrm{mem}$ achieved a BER of $0.03(3)$ at $N_v = 12$, $\mathrm{TBP}_\mathrm{mem} = 6.4$.

\begin{figure}[t]
    \centering
    \includegraphics[width=\linewidth]{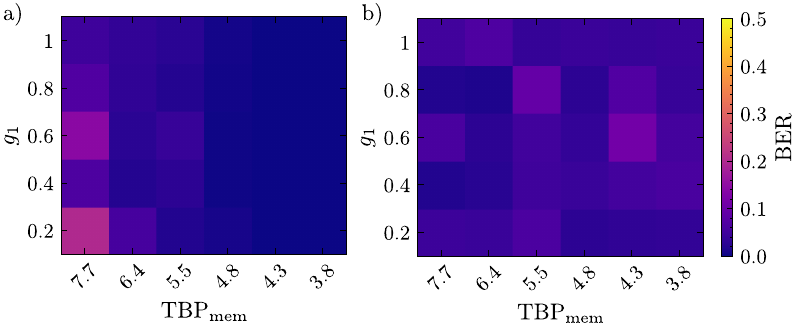}
  \caption{XOR performance (BER) as a function of input scaling $g_1$ and $\mathrm{TBP}_\mathrm{mem}$ for $N_v = 12$, $T_1 = 230~\mu$s. (a)~Simulation: BER$= 0$ across all $g_1$ for $\mathrm{TBP}_\mathrm{mem} \geq 4.8$. (b)~Experiment: minimum BER$= 0.02(1)$ at $g_1 = 0.6$, $\mathrm{TBP}_\mathrm{mem} = 6.4$.}
  \label{fig:2XOR_g1_dt}
\end{figure}

Figure~\ref{fig:2XOR_g1_dt} presents the XOR BER as a function of $g_1$ at $N_v = 12$. The range of $g_1$ is reduced to $[0.2,1]$, because simulations such as those shown in Figure \ref{fig:2_KR_g1} did not suggest any significant gain in the $g_1 > 1$ range. Here we achieved the minimum experimental BER of $0.02(1)$ at $g_1 = 0.6$, $\mathrm{TBP}_\mathrm{mem} = 6.4$. Crucially, varying $g_1$ across the full range from linear ($g_1 = 0.2$) to saturated ($g_1 = 1.0$) operation produces negligible change in performance, corroborating the finding that the atomic saturation nonlinearity contributes weakly to computation in this system. Simulation predicts zero BER for all values of $g_1$ when $\mathrm{TBP}_\mathrm{mem} \geq 4.8$, indicating that the nonlinearity provided by the network topology is sufficient to solve XOR once adequate memory depth is established. The discrepancy between simulated and experimental BER is attributed to measurement noise.

\subsection{Discussion}

The results reveal a principal limitation of the current implementation: the low memory lifetime is insufficient for deep temporal processing. The diffusion-limited lifetime $\tau_\mathrm{mem} \approx 1.3~$ms is not reached in practice; the measured $T_1 = 230~\mu$s for standard pump parameters limits the logical TBP to approximately 2.6. Increasing $N_v$ to improve network complexity proportionally consumes the available memory budget (Eq.~\ref{eq:tbp}), creating a fundamental complexity-memory trade-off. This trade-off is structurally distinct from delay-line reservoirs, where the virtual node count can be increased by faster sampling without extending the total injection time. 
Moreover, while the system’s short temporal dynamics and nonlinear projection suggest potential for use as an extreme learning machine, the single-input reservoir architecture prevents the use of multi-feature inputs and therefore limits its applicability as an ELM.

Several strategies may mitigate the limited relaxation lifetime of the system. These include implementing active re-pumping to refresh atomic polarization or employing electronic pre-processing to augment the input with delayed copies, thereby shifting the memory requirements to the pre-processing layer \cite{jaurigueReservoirComputingDelayed2021, piccoEfficientOptimisationPhysical2025}. Future work could also explore high-speed addressing via AOMs and fast writing of coherent memories to accelerate read/write cycles \cite{messnerMultiplexedRandomaccessOptical2023}, or parallelized operations to update multiple virtual nodes within a single time step. Finally, transitioning to telecom-compatible hardware, such as Erbium-doped fiber memories, offers a promising alternative due to their demonstrated capacity for highly multiplexed operation \cite{kamelErbiumDopedFibreQuantum2025, saglamyurekQuantumStorageEntangled2015}.

Despite these constraints, this work establishes that warm \Cs\ atomic vapor is a viable substrate for free-space optical RAM and demonstrates its integration into a reservoir computer. The system successfully performs XOR with BER~$= 0.02(1)$, and the ORAM supports accumulative in-memory writing in the linear regime.

\section{Conclusion}

We have demonstrated an optically writable spatially multiplexed ORAM based on warm \Cs\ atomic vapor and its application as a physical substrate for a reservoir computer. The memory operates through optical pumping of the $D_1$ hyperfine transition, with eight addressable spatial rails defined by acousto-optic deflection. Characterization yielded a $1/e$ lifetime of $T_1 = 230~\mu$s and a maximum logical time-bandwidth product of $\mathrm{TBP}_\mathrm{logic} = 2.6$.

The reservoir benchmarking showed a maximum linear memory capacity of LMC$= 1.0\pm0.2$ and a kernel rank of KR$= 8.8\pm0.4$, both primarily governed by the network topology rather than the nonlinearity arising from atomic saturation. The system successfully solved the XOR classification task with a minimum BER of $0.02\pm0.01$. Performance is fundamentally limited by the interplay between memory lifetime and the temporal multiplexing overhead, which restricts the logical TBP and confines the reservoir to an ELM-like operating regime for tasks requiring multi-step memory. Future research will focus on optimizing read/write protocols for storage rails and evaluating the suitability of high-speed memory architectures, such as those based on electromagnetically induced transparency (EIT).
\begin{backmatter}
\bmsection{Funding}
This research was funded by the Deutsche Forschungsgemeinschaft (DFG),  Grant No. LU 1729/3-1 and WO 2218/5-1 (Projektnummer 445183921). E.R. acknowledges funding through the Helmholtz Einstein International Berlin Research School in Data Science (HEIBRiDS).
\bmsection{Acknowledgment}

\bmsection{Disclosures}
The authors declare no conflicts of interest.

\bmsection{Data availability} Data underlying the results presented in this paper may be obtained from the authors upon reasonable request.

\end{backmatter}

\bibliography{references} 

\end{document}